\documentclass[12pt,a4paper]{article}
\usepackage{amsmath,amsfonts}
\usepackage{braket}
\usepackage[nosort]{cite}
\newlength{\extraspace}
\newlength{\extraspaces}
 \catcode`\@=11
\def\numberbysection{\@addtoreset{equation}{section}
\def\theequation{\arabic{section}.\arabic{equation}}}
\newcommand{\newsection}[1]{
\vspace{7mm}
\pagebreak[3]
\addtocounter{section}{1}
\setcounter{equation}{0}
\setcounter{subsection}{0}
\setcounter{footnote}{0}
\begin{center}
{\large {\bf \thesection. #1}}
\end{center}
\nopagebreak
\medskip
\nopagebreak
\hspace{3mm}}
\newcommand{\nonu}{\nonumber \\[.5mm]}

\newcommand{\tr}{\, \textrm{tr}}

\newcommand{\VEV}[1]{\left\langle {#1} \right\rangle}

\numberbysection
\begin{document}
\addtolength{\baselineskip}{.7mm}

\thispagestyle{empty}
\begin{flushright}
STUPP--19--236 \\
May, 2019
\end{flushright}
\vspace{20mm}
\begin{center}
{\Large \textbf{Linearized Field Equations of Gauge Fields \\[2mm]
from the Entanglement First Law}} \\[20mm]
\textsc{Kenta Hasegawa}\footnote{
\texttt{e-mail: hasegawa@krishna.th.phy.saitama-u.ac.jp}}
\hspace{1mm} and \hspace{1mm}
\textsc{Yoshiaki Tanii}\footnote{
\texttt{e-mail: tanii@phy.saitama-u.ac.jp}} \\[7mm]
\textit{Division of Material Science \\ 
Graduate School of Science and Engineering \\
Saitama University, Saitama 338-8570, Japan} \\[20mm]
\textbf{Abstract}\\[7mm]
{\parbox{13cm}{\hspace{5mm}
In the context of the AdS/CFT correspondence linearized field 
equations of vector and antisymmetric tensor gauge fields 
around an AdS background are obtained from the entanglement 
first law of CFTs. The holographic charged entanglement entropy 
contains a term depending on the gauge field in addition 
to the Ryu--Takayanagi formula. 
}}
\end{center}
\vfill
\newpage
\setcounter{section}{0}
\setcounter{equation}{0}
\numberbysection
%
%
\newsection{Introduction}
The idea of entanglement has been discussed in the context of 
the AdS/CFT correspondence 
\cite{Maldacena:1997re,Gubser:1998bc,Witten:1998qj}, which 
relates a conformal field theory (CFT) in Minkowski spacetime 
and a gravitational theory in higher dimensional anti de Sitter 
(AdS) spacetime. 
In particular, Ryu and Takayanagi proposed a direct connection 
between the entanglement entropy of a CFT to a dual bulk 
geometry \cite{Ryu:2006bv,Ryu:2006ef}, which was generalized 
to a covariant form in \cite{Hubeny:2007xt}. 
(For reviews see \cite{Rangamani:2016dms,Nishioka:2018khk}.)

Recently, the entanglement entropy has been used to understand 
how the bulk gravitational dynamics is obtained from a CFT 
\cite{Nozaki:2013vta,Allahbakhshi:2013rda,Lashkari:2013koa,
Bhattacharya:2013bna,Faulkner:2013ica,Swingle:2014uza,
Kastor:2014dra,Lin:2014hva,Jacobson:2015hqa,Speranza:2016jwt,
Kastor:2016bph,Caceres:2016xjz,Mosk:2016elb,Czech:2016tqr,
Faulkner:2017tkh,Paul:2018spp}. 
In \cite{Lashkari:2013koa,Faulkner:2013ica} the linearized field 
equation of the gravitational field around AdS spacetime was 
derived from a property of the entanglement entropy of the CFT. 
The entanglement entropy satisfies 
the entanglement first law \cite{Blanco:2013joa}, 
which relates a variation of the entanglement entropy and that 
of the expectation value of the modular Hamiltonian. 
By rewriting this relation in terms of the bulk gravitational 
field by the AdS/CFT correspondence one obtains a constraint 
on the gravitational field, which turns out to be the linearized 
field equation.

The purpose of this paper is to extend the result of 
\cite{Lashkari:2013koa,Faulkner:2013ica} and show that linearized 
field equations of vector and antisymmetric tensor gauge fields 
also can be derived from the entanglement first law. 
Since gravitational theories dual to CFTs, such as supergravity 
and superstring theories, contain fields other 
than the gravitational field, it is natural to consider 
a possibility to derive their field equations from the 
entanglement first law.

To derive the linearized field equation of a vector field 
we consider a CFT with a conserved $U(1)$ current $J^\mu$. 
In the AdS/CFT correspondence the boundary value of the bulk 
vector field plays a role of a source for this current. 
Using the charge of this current we can define the charged 
entanglement entropy 
\cite{Lewkowycz:2013nqa,Wong:2013gua,Belin:2013uta}, 
which satisfies the entanglement first law similar to the 
first law of thermodynamics for a grand canonical ensemble.  
By rewriting the first law in terms of the bulk fields we obtain 
the linearized field equations of the vector field as well 
as the gravitational field. 
To rewrite the first law in terms of the bulk fields we follow 
the approach of \cite{Faulkner:2013ica}, which uses the Noether 
charge of local symmetries of the bulk theory. 
In \cite{Faulkner:2013ica} the Noether charge for a coordinate 
transformation by a Killing vector was used. 
In that case the field equation of only the gravitational field 
was obtained. Even when matter fields are present in the bulk theory, 
they contribute to the first law only at higher orders in perturbations 
and their linearized field equations are not obtained. 
In our work we consider a $U(1)$ gauge transformation of the vector 
field which preserves the background configuration in addition to 
the coordinate transformation. This allows us to obtain the 
linearized field equation of the vector field from the entanglement 
first law. In this calculation 
we find that the entanglement entropy expressed 
by the bulk fields (\ref{cee2}) has an extra term depending 
on the vector field in addition to the Ryu--Takayanagi formula 
proportional to the area of the extremal surface.

The discussion for a vector field can be generalized to 
the case of an antisymmetric tensor field. 
By considering a CFT with conserved antisymmetric tensor current 
$J^{\mu_1\cdots\mu_n}$ we obtain the linearized field equation 
of an $n$-th rank antisymmetric tensor field 
from the entanglement first law. The charged entanglement entropy 
(\ref{cee3}) contains a term depending on the antisymmetric 
tensor field in addition to the Ryu--Takayanagi formula.

The organization of this paper is as follows. 
In the next section we discuss the charged entanglement entropy 
of a CFT and the entanglement first law. 
In section 3 we consider a bulk theory consisting of 
a gravitational field and a vector field. 
We rewrite the entanglement first law in terms of the bulk fields, 
from which the linearized field equations are derived. 
In section 4 the linearized field equation of an antisymmetric 
tensor field is derived from the entanglement first law 
in a similar way. We conclude in section 5. 
In appendix A we discuss the holographic renormalization of 
an antisymmetric tensor field and derive a formula for the 
one-point function of the CFT current, which we use in the text. 
In appendix B we discuss another derivation of that formula 
without using the holographic renormalization calculation.

%
\newsection{Charged entanglement entropy}
We consider a CFT in $d$-dimensional flat Minkowski spacetime, 
which has an energy-momentum tensor $T_{\mu\nu}$ 
and a $U(1)$ current $J^\mu$ satisfying
\begin{equation}
\partial_\mu T^{\mu\nu} = 0, \qquad
T_\mu{}^\mu = 0, \qquad
\partial_\mu J^\mu = 0. 
\label{conservation}
\end{equation}
We assume that this CFT is dual to a $(d+1)$-dimensional 
classical gravitational theory as discussed in the next section. 
In order to define the entanglement entropy in the CFT we choose 
a CFT state $\ket{\psi}$ and a region $B$ on a time slice $t=t_0$. 
As in \cite{Casini:2011kv,Lashkari:2013koa,Faulkner:2013ica} 
we consider the case in which $B$ is a ball of radius $R$ centered 
at a spatial point $x_0^i$ ($i=1,2,\cdots,d-1$). 
The state of the region $B$ is described by the reduced 
density matrix 
\begin{equation}
\rho_B = \tr_{\bar{B}} \, \rho_{\textrm{total}}
= \frac{e^{-H_B}}{\tr_B (e^{-H_B})},
\label{rdm}
\end{equation}
where $\rho_{\textrm{total}} = \ket{\psi}\bra{\psi}$ is the pure 
density matrix of the full system corresponding to the state 
$\ket{\psi}$, and $\tr_{\bar{B} }$ means tracing over states 
in $\bar{B}$, the complement of $B$ on the time slice $t=t_0$. 
The density matrix $\rho_B$ can be expressed by the operator $H_B$ 
called the modular Hamiltonian as in (\ref{rdm}). 
The entanglement entropy is defined as the von Neumann entropy of 
this reduced density matrix 
$S_B = - \tr_B \left( \rho_B \log \rho_B \right)$.
Using the charge of the $U(1)$ current $J^\mu$
we can also define the charged entanglement entropy 
\cite{Lewkowycz:2013nqa, Wong:2013gua, Belin:2013uta}. 
We first introduce a new density matrix 
\begin{equation}
\rho_B(\mu) = \frac{e^{-H_B + \mu Q_B}}{\tr_B(e^{-H_B + \mu Q_B})},
\label{rdm2}
\end{equation}
where 
\begin{equation}
Q_B = \int_B d^{d-1}x \, J^0
\label{bcharge}
\end{equation}
is the charge operator in $B$ and $\mu$ is a constant. 
Then, the charged entanglement entropy is defined as 
\begin{equation}
S_B(\mu) = - \tr_B \left[ \rho_B(\mu) \log \rho_B(\mu) \right] .
\label{cee}
\end{equation}

Now, consider an infinitesimal variation of the CFT state 
$\ket{\psi} \rightarrow \ket{\psi} + \ket{\delta\psi}$, 
which induces a variation of $\rho_B(\mu)$. 
The first order variation of the charged entanglement entropy 
(\ref{cee}) then gives the first law of the entanglement 
\begin{equation}
\delta S_B(\mu) = \delta \VEV{H_B} - \mu \, \delta \VEV{Q_B},
\label{firstlaw}
\end{equation}
where $H_B$ is the unperturbed modular Hamiltonian, and 
the expectation value of an operator $O$ in $B$ is defined as 
\begin{equation}
\VEV{O} \equiv \tr_B\left( \rho_B(\mu) \, O \right) 
= \bra{\psi} e^{\mu Q_B} O \ket{\psi}.
\label{cftvev}
\end{equation}
Here, we have assumed that $Q_B$ commutes with $H_B$. 
This is indeed the case in our setup as we will discuss 
in the next paragraph. 
The first law (\ref{firstlaw}) resembles the first law of 
thermodynamics for a grand canonical ensemble. 
The constant $\mu$ corresponds to a chemical potential 
in thermodynamics. 
In the following we consider the case in which the unperturbed 
state $\ket{\psi}$ is the CFT vacuum $\ket{0}$. 
We will show in the next section that the first law 
(\ref{firstlaw}) leads to linearized field equations of 
bulk gravitational and vector fields.

The modular Hamiltonian $H_B$ is known when the CFT state is 
the vacuum $\ket{\psi}=\ket{0}$ and $B$ is the ball-shaped 
region of radius $R$. It is given by 
\begin{equation}
H_B = 2\pi \int_{B}d^{d-1}x \, \frac{R^2 - (x^i-x^i_0)^2}{2R} \, 
T_{00} , 
\label{modularh2}
\end{equation}
where $T_{00}$ is the $00$ component of the energy-momentum tensor 
$T_{\mu\nu}$. This formula was obtained in \cite{Casini:2011kv} 
as follows. 
By a conformal transformation the causal development of the 
region $B$ in $d$-dimensional Minkowski spacetime is mapped 
to a hyperbolic cylinder $\mathbb{R} \times H^{d-1}$, 
where $H^{d-1}$ is a $(d-1)$-dimensional hyperboloid 
of curvature radius $R$, and $\mathbb{R}$ represents time. 
The modular Hamiltonian $H_B$ in (\ref{modularh2}) and the 
charge operator $Q_B$ in (\ref{bcharge}) are obtained from 
the Hamiltonian $H$ and the charge operator $Q$ in the hyperbolic 
cylinder as 
\begin{equation}
H_B = 2\pi R \, U^{-1} H U, \qquad
Q_B = U^{-1} Q U, 
\label{conformaltransformation}
\end{equation}
where $U$ is the unitary operator which implements 
the conformal transformation. 
Since $Q$ is a conserved charge defined on the entire 
space $H^{d-1}$, it commutes with the Hamiltonian $H$. 
As a consequence $H_B$ and $Q_B$ in (\ref{conformaltransformation}) 
also commute each other as we mentioned below (\ref{cftvev}). 
The density matrix (\ref{rdm2}) is then related to 
the thermal density matrix in the hyperbolic cylinder with 
temperature $T=(2 \pi R)^{-1}$ and chemical potential $\mu$ as 
\begin{equation}
\rho_B(\mu) 
= U^{-1} \frac{e^{-H/T + \mu Q}}{\tr(e^{-H/T + \mu Q})} U. 
\end{equation}
Therefore, the charged entanglement entropy (\ref{cee}) is equal 
to the thermal entropy of the CFT in the hyperbolic cylinder. 
By the AdS/CFT correspondence this thermal entropy can be 
calculated as the entropy of a black hole with a hyperbolic 
horizon in the bulk. This was done in \cite{Faulkner:2013ica} for 
the case $\mu=0$ by using Wald's formula of the horizon entropy 
\cite{Wald:1993nt,Iyer:1994ys}. 
We will generalize it to the case $\mu \not= 0$ in the next section.

%
\newsection{Linearized field equations}
In this section we first rewrite each side of the entanglement 
first law (\ref{firstlaw}) in terms of bulk fields by the 
AdS/CFT correspondence. 
Then, the first law will require that perturbations of the bulk 
fields corresponding to the variation of the CFT state
$\ket{\delta\psi}$ should satisfy linearized field equations. 
This was shown in \cite{Lashkari:2013koa, Faulkner:2013ica} 
for the bulk gravitational field in the case $\mu=0$. 
We will generalize that result by introducing a bulk vector field 
corresponding to the CFT current $J^\mu$. 
The entanglement first law will then require that perturbations of 
the vector field as well as the gravitational field should satisfy 
linearized field equations.

The Lagrangian for the gravitational field $g_{ab}$ and a $U(1)$ 
gauge field $A_a$ in $(d+1)$-dimensional bulk spacetime 
is\footnote{We use $a,b, \cdots=0,1,\cdots, d$ for ($d+1$)-dimensional 
coordinate indices.} 
\begin{equation}\label{lagrangian}
\mathcal{L} = \frac{1}{16\pi} 
\sqrt{-g} \left( R + \frac{d(d-1)}{l^2} \right) 
- \frac{1}{4} \sqrt{-g} F_{ab} F^{ab},
\end{equation}
where $l$ is a constant characterizing the cosmological constant. 
We have chosen the gravitational constant as $G=1$ for simplicity. 
Under general variations of the fields the Lagrangian changes as 
\begin{equation}
\delta \mathcal{L} 
= \frac{1}{16\pi} \sqrt{-g} \left( \delta g^{ab} E_{ab} 
+ 16\pi \delta A_a E^a + \nabla_a v^a \right), 
\label{generalvariation}
\end{equation}
where 
\begin{align}
E_{ab} &= R_{ab} - \frac{1}{2} g_{ab} R 
- \frac{d(d-1)}{2l^2} \, g_{ab} - 8\pi T^{\rm bulk}_{ab}, \nonu
E^a &= \nabla_b F^{ba}, \nonu
T^{\rm bulk}_{ab} &= F_{ac} F_b{}^c - \frac{1}{4} g_{ab} F_{cd} F^{cd}, \nonu
v_a &= \nabla^b \delta g_{ab} - g^{cd} \nabla_a \delta g_{cd} 
- 16\pi F_a{}^b \delta A_b.
\label{fieldeqs}
\end{align}
$E_{ab}=0$ and $E^a=0$ are the field equations of $g_{ab}$ and 
$A_a$ respectively with $T^{\rm bulk}_{ab}$ being the 
energy-momentum tensor of the vector field. 
Under general coordinate transformations and $U(1)$ gauge transformations
\begin{align}
\delta_\xi g_{ab} &= \nabla_a \xi_b + \nabla_b \xi_a, \nonu
\delta_\xi A_a &= \xi^b \partial_b A_a + \partial_a \xi^b A_b 
+ \partial_a \xi,
\label{localtransf}
\end{align}
the Lagrangian is invariant up to a total divergence 
\begin{equation}
\delta_\xi \mathcal{L} = \partial_a \left( \xi^a \mathcal{L} \right) .
\label{localtransfl}
\end{equation}

To find a bulk representation of each side of (\ref{firstlaw}) 
we first consider the Noether current corresponding to the local 
symmetry transformations (\ref{localtransf}) following 
\cite{Wald:1993nt,Iyer:1994ys,Gao:2001ut,Gao:2003ys}. 
The Noether current is 
\begin{equation}
J^a = \frac{1}{16\pi} \, \bar{v}^a 
- \xi^a \frac{\mathcal{L}}{\sqrt{-g}} , 
\end{equation}
where $\bar{v}^a$ is given by $v^a$ in (\ref{fieldeqs}) 
with $\delta g_{ab}$, $\delta A_a$ replaced 
by $\delta_\xi g_{ab}$, $\delta_\xi A_a$ 
in (\ref{localtransf}).  By (\ref{generalvariation}) 
and (\ref{localtransfl}) $J^a$ satisfies 
\begin{equation}
\nabla_a J^a = - \frac{1}{16\pi} \delta_\xi g^{ab} E_{ab} 
- \delta_\xi A_a E^a 
\label{onshellcon}
\end{equation}
and therefore is divergence free on-shell, i.e., when the field 
equations $E_{ab}=0$, $E^a=0$ are satisfied. 
As discussed in \cite{Iyer:1995kg,Barnich:2001jy,Avery:2015rga} 
we can construct a new current $\tilde{J}^a$ which coincides 
with $J^a$ on-shell and is divergence free off-shell. 
Indeed, the right-hand side of (\ref{onshellcon}) can be 
written as a divergence $\nabla_a S^a$, where  
\begin{equation}
S^a = \frac{1}{8\pi} \xi_b E^{ab} 
- \left(\xi \cdot A + \xi \right) E^a, 
\end{equation}
and the new current $\tilde{J}^a = J^a - S^a$ satisfies 
$\nabla_a \tilde{J}^a = 0$ off-shell. Since $S^a = 0$ on-shell, 
$\tilde{J}^a$ coincides with $J^a$ on-shell. 
In terms of differential forms $\nabla_a \tilde{J}^a = 0$ 
can be written as\footnote{We use boldface letters to denote 
differential forms.}  
\begin{equation}
d \tilde{\boldsymbol{J}} = 0, \qquad
\tilde{\boldsymbol{J}} 
= \tilde{J}^b \, \boldsymbol{\epsilon}_b, 
\label{closed} 
\end{equation}
and we find that $\tilde{\boldsymbol{J}}$ is an exact form 
$\tilde{\boldsymbol{J}} = d \boldsymbol{Q}$, where 
\begin{equation}
\boldsymbol{Q} = \left[ - \frac{1}{16\pi} \nabla^{b} \xi^{c} 
- \frac{1}{2} \left( \xi \cdot A + \xi \right) F^{bc} \right]
\boldsymbol{\epsilon}_{bc}. 
\label{exact}
\end{equation}
In (\ref{closed}), (\ref{exact}) we used the notation 
\begin{align}
\boldsymbol{\epsilon}_b &= \frac{1}{d!} \, \epsilon_{b a_1 \cdots a_d} 
\, dx^{a_1} \wedge \cdots \wedge dx^{a_d}, \nonu
\boldsymbol{\epsilon}_{bc} &= \frac{1}{(d-1)!} \, 
\epsilon_{bc a_1 \cdots a_{d-1}} 
\, dx^{a_1} \wedge \cdots \wedge dx^{a_{d-1}}, 
\end{align}
where $\epsilon_{a_1 \cdots a_{d+1}}$ is the totally antisymmetric 
tensor with non-vanishing components $\pm \sqrt{-g}$.

We then split the fields as 
$g_{ab} \rightarrow g_{ab} + \delta g_{ab}$, 
$A_a \rightarrow A_a + \delta A_a$, 
where $g_{ab}$, $A_a$ are background fields satisfying the field 
equations $E_{ab}=0$ and $E^a=0$, and $\delta g_{ab}$, $\delta A_a$ 
are small perturbations around the background. 
In the setting of the CFT in the previous section 
the background corresponds to the vacuum $\ket{\psi}=\ket{0}$ 
and the perturbations correspond to an infinitesimal variation 
of the state $\ket{\delta\psi}$. 
The background corresponding to the CFT vacuum is the AdS metric 
and a gauge field with vanishing field strength: 
\begin{equation}
g_{ab} dx^a dx^b = \frac{l^2}{z^2} \left( dz^2 
+ \eta_{\mu\nu} dx^\mu dx^\nu \right), \qquad
F_{ab} = 0.
\label{background}
\end{equation}
Here, the coordinate $z$ takes values $z > 0$, and 
the AdS boundary at infinity $z=0$ corresponds to 
the Minkowski spacetime with coordinates $x^\mu$ 
($\mu=0,1,\cdots,d-1$), in which the CFT is defined. 
The factor $e^{\mu Q_B}$ in the expectation value in (\ref{cftvev}) 
means that the vector field $A_a$ has a non-vanishing 
background proportional to $\mu$ for $z \rightarrow 0$ 
\cite{Lewkowycz:2013nqa,Wong:2013gua,Belin:2013uta}.

The background (\ref{background}) is invariant under the 
transformations (\ref{localtransf}) when $\xi^a$ is a Killing 
vector of AdS. As in \cite{Faulkner:2013ica} we use the Killing 
vector 
\begin{equation}
\xi^a \partial_a =  - \frac{2\pi}{R} (t-t_0) \left[ 
z \partial_z + (x^i-x^i_0) \partial_i \right] 
+ \frac{\pi}{R} \left[ R^2 - z^2 - (t-t_0)^2 
- (x^i-x_0^i)^2 \right] \partial_t, 
\label{bulkkilling}
\end{equation}
which approaches to the $d$-dimensional Killing vector corresponding 
to time translation of the hyperbolic cylinder at the boundary $z=0$. 
When we use the gauge field $A_a$ itself instead of the field 
strength $F_{ab}$, we also need to consider a compensating $U(1)$ 
gauge transformation. 
To find it we note that the transformation of the gauge field 
in (\ref{localtransf}) can be rewritten as 
\begin{equation}
\delta_\xi A_a = \xi^b F_{ba} + \partial_a 
\left( \xi \cdot A + \xi \right). 
\end{equation}
Therefore, the gauge field $A_a$ with $F_{ab}=0$ is invariant 
when we choose the $U(1)$ gauge transformation parameter as 
\begin{equation}
\xi = \mu' - \xi \cdot A, 
\label{gaugeparameter}
\end{equation}
where $\mu'$ is an arbitrary constant. Later we will choose 
this constant as $\mu'=\mu$, where $\mu$ is the chemical 
potential appearing in the entanglement first law (\ref{firstlaw}).

We can now construct a $(d-1)$-form $\boldsymbol{\chi}$ from 
the bulk fields which gives each side of the entanglement 
first law (\ref{firstlaw})  as 
\begin{equation}
\int_{B} \boldsymbol{\chi} = \delta \VEV{H_B} - \mu \delta \VEV{Q_B}, 
\qquad \int_{\tilde{B}} \boldsymbol{\chi} = \delta S_B(\mu). 
\label{eq01}
\end{equation}
Here, $B$ is the ball-shaped region of radius $R$  
centered at $x^{i}=x_{0}^{i}$ on a time slice $t=t_0$ 
in $d$-dimensional Minkowski spacetime at the boundary $z=0$: 
\begin{equation}
B=\{(z, x^\mu) \, | \ t=t_0, \ z=0,\ (x^i -x^i_0)^2 \leq R^2 \}. 
\end{equation}
$\tilde{B}$ is the extremal surface in the bulk which 
has the same boundary as $B$ and is homologous to $B$. 
$\tilde{B}$ has the extremal area with respect to the background 
metric (\ref{background}) and turns out to be a hemisphere 
\begin{equation}
\tilde{B}=\{ (z, x^\mu) \, | \ t=t_0, \ (x^i -x^i_0)^2 +z^2 =R^2 \}. 
\end{equation}
The desired $(d-1)$-form is given by 
\begin{align}
\boldsymbol{\chi} = \delta \boldsymbol{Q} 
- \frac{1}{16\pi} v^b \xi^c \boldsymbol{\epsilon}_{bc}, 
\label{chiform}
\end{align}
where $\boldsymbol{Q}$ and $v^a$ are given in (\ref{exact}) and 
(\ref{fieldeqs}). The transformation parameters $\xi^a$ and 
$\xi$ are those in (\ref{bulkkilling}) and (\ref{gaugeparameter}) 
with $\mu'=\mu$. 
$\delta$ denotes variations of the fields $g_{ab}$, 
$A_a$ with the transformation parameters $\xi^a$, $\xi$ fixed. 
In the following we will show that this $\boldsymbol{\chi}$ indeed 
satisfies (\ref{eq01}). 
We impose a gauge condition on the perturbations of the fields as 
\begin{equation}
\delta g_{zz} = 0, \quad
\delta g_{z\mu} = 0, \quad
\delta A_z = 0. 
\end{equation}

First, let us examine the integral over $B$ of the first equation 
in (\ref{eq01}). Using (\ref{background}), (\ref{bulkkilling}), 
(\ref{gaugeparameter}) we find 
\begin{equation}
\int_B \boldsymbol{\chi} 
= \int d^{d-1}x \left. \left[ \frac{d l^{d-3}}{16R} 
\left\{ R^2 - (x^i-x^i_0)^2 \right\} \delta g^{(d-2)}_{00} 
+ (d-2) \, l^{d-3} \mu \, \delta A^{(d-2)}_0 \right] 
\right|_{z=0},
\label{chib}
\end{equation}
where we have put 
\begin{equation}
\delta g_{\mu\nu}(x,z)=z^{d-2} \delta g^{(d-2)}_{\mu\nu}(x), \qquad
\delta A_\mu(x,z) = z^{d-2} \delta A^{(d-2)}_\mu(x)
\label{deltaga}
\end{equation}
for $z \rightarrow 0$ so that (\ref{chib}) takes a finite value. 
By the holographic renormalization 
\cite{deHaro:2000vlm,Skenderis:2002wp} 
$\delta g^{(d-2)}_{\mu\nu}$ and $\delta A^{(d-2)}_\mu$ 
are related to the one-point functions of the energy-momentum 
tensor $T_{\mu \nu}$ and the current $J_{\mu}$ of the CFT as 
\begin{equation}
\delta \VEV{T_{\mu\nu}} 
= \frac{d l^{d-3}}{16\pi} \delta g^{(d-2)}_{\mu\nu}, \qquad
\delta \VEV{J_\mu} 
= (d-2) \, l^{d-3} \delta A^{(d-2)}_\mu. 
\label{hren}
\end{equation}
The first relation in (\ref{hren}) was already used in 
\cite{Lashkari:2013koa, Faulkner:2013ica}. 
The second relation is obtained in appendix A. 
Substituting (\ref{hren}) into (\ref{chib}) we obtain 
\begin{equation}
\int_B \boldsymbol{\chi} = \int_{B}d^{d-1}x \left[ 2\pi \frac{R^2 
- (x^i-x^i_{0})^2}{2R} \delta \VEV{T_{00}} 
- \mu \delta \VEV{J^0} \right]. 
\end{equation}
Using (\ref{modularh2}), (\ref{bcharge}) we find that the 
first equation in (\ref{eq01}) is indeed satisfied.

Next, let us examine the integral over $\tilde{B}$ of the 
second equation in (\ref{eq01}). 
Using (\ref{background}), (\ref{bulkkilling}), (\ref{gaugeparameter}) 
we find 
\begin{align}
\int_{\tilde{B}} \boldsymbol{\chi} 
&= \int_B d^{d-1}x \left[ \frac{l^{d-3}}{8R \, z^{d-2}} 
\left( R^2 \delta^{ij} - (x-x_0)^i (x-x_0)^j \right) \delta g_{ij}
\right. \nonu
& \quad \left. \left. 
+ \frac{l^{d-3}}{z^{d-3}} \, \mu \left( 
\delta F_{z0} + \frac{(x-x_0)^i}{z} \delta F_{i0} \right)
\right]\right|_{z=\sqrt{R^2-(x^i-x^i_0)^2}}. 
\end{align}
As was shown in \cite{Lashkari:2013koa, Faulkner:2013ica} 
the first term is the variation of the Ryu--Takayanagi formula 
proportional to the area $A$ of the extremal surface for 
the metric $g_{ab}+\delta g_{ab}$. 
The second term is an additional contribution depending on 
the gauge field. If we assume that the charged 
entanglement entropy is given by\footnote{The entanglement 
entropy of this form was previously used as an order parameter 
that distinguishes various phases of field theories 
\cite{Hartnoll:2012ux}. We thank Juan F. Pedraza for informing 
us of this work.} 
\begin{equation}
S_B(\mu) = \frac{1}{4} A 
- \mu \int_{\tilde{B}} * \boldsymbol{F},
\label{cee2}
\end{equation}
where $*\boldsymbol{F}$ is the Hodge dual of $\boldsymbol{F}$, 
then the second equation in (\ref{eq01}) is satisfied.

Once we accept (\ref{cee2}) as the formula for the charged 
entanglement entropy, we can use it to derive the relations 
in (\ref{hren}) between the expectation values of the CFT 
operators and the asymptotic values of the fields 
without using the holographic renormalization calculation. 
In \cite{Faulkner:2013ica} the first relation in (\ref{hren}) 
for the energy-momentum tensor was indeed derived from the 
entanglement first law (\ref{firstlaw}) with $\mu=0$ 
and the Ryu--Takayanagi formula by considering a small size 
limit of the ball-shaped region $B$. 
Similarly, the second relation in (\ref{hren}) for the current 
can be derived from the $\mu$-dependent terms of (\ref{firstlaw}) 
in the same limit. 
This derivation is discussed in appendix B.

Thus, we have shown that $\boldsymbol{\chi}$ in (\ref{chiform}) 
satisfies (\ref{eq01}) assuming that the charged entanglement 
entropy is given by (\ref{cee2}). 
The entanglement first law (\ref{firstlaw}) then requires 
\begin{equation}
0 = \int_{\tilde{B}} \boldsymbol{\chi} - \int_{B} \boldsymbol{\chi}
= \int_{\Sigma} d \boldsymbol{\chi}, 
\label{efl2}
\end{equation}
where $\Sigma$ is the region enclosed by $\tilde{B}$ and $B$ 
on the time slice $t=t_0$ satisfying $\partial \Sigma=\tilde{B}-B$.
The exterior derivative of $\boldsymbol{\chi}$ in (\ref{chiform}) 
is found to be 
\begin{equation}
d \boldsymbol{\chi} 
= \left( - \frac{1}{8\pi} \delta E^{ab} \xi_b 
+ \mu \delta E^a \right) \boldsymbol{\epsilon}_a,
\label{dchi}
\end{equation}
where $\delta E_{ab}$ and $\delta E^a$ are variations of 
$E_{ab}$ and $E^a$ in (\ref{fieldeqs}): 
\begin{align}
\delta E_{ab} 
&= \frac{1}{2} \left( \nabla^c \nabla_a \delta g_{bc} 
+ \nabla^c \nabla_b \delta g_{ac} - \nabla^2 \delta g_{ab} 
- g^{cd} \nabla_a \nabla_b \delta g_{cd} \right) \nonu
& \quad - \frac{1}{2} g_{ab} \left( \nabla^c \nabla^d \delta g_{cd} 
- g^{cd} \nabla^2 \delta g_{cd} \right) 
+ \frac{d}{l^2} \, \delta g_{ab}
- \frac{d}{2l^2} \, g_{ab} g^{cd} \delta g_{cd}, \nonu
\delta E^a &= g^{ab} \nabla^c \delta F_{cb}.
\label{lfeq}
\end{align}
$\delta E_{ab}=0$ and $\delta E^a=0$ are the 
linearized field equations around the background 
(\ref{background}).\footnote{The on-shell closed form $\boldsymbol{\chi}$
may be understood as a calibration \cite{Bakhmatov:2017ihw}. 
We thank Eoin \'O Colg\'ain for pointing it out to us.}
Substituting (\ref{dchi}) into (\ref{efl2}) we obtain 
\begin{equation}
\int_{\Sigma} dz d^{d-1}x \left( g^{00} \delta E_{00} \xi^0 
- 8\pi \mu \delta E^0 \right) = 0, 
\end{equation}
where we have used the fact that only the time components of 
$\xi^a$ and $\boldsymbol{\epsilon}_a$ are non-vanishing on $\Sigma$. 
By requiring that this condition holds for arbitrary $R$, $x_0^i$, 
$t_0$ and $\mu$ we obtain the local conditions 
$\delta E_{00} = 0$, $\delta E^0 = 0$ 
(See appendix A of \cite{Faulkner:2013ica}).  
Moreover, requiring it for any frame of reference we obtain 
$\delta E_{\mu\nu} = 0$, $\delta E^\mu = 0$ 
($\mu,\nu=0,1,\cdots,d-1$).

In \cite{Faulkner:2013ica} it was shown that the remaining 
gravitational equations $\delta E_{z\mu}=0$ and $\delta E_{zz}=0$ 
are obtained from $\delta E_{\mu\nu}=0$ and the 
tracelessness and the conservation of the CFT energy-momentum 
tensor $T_{\mu\nu}$ in (\ref{conservation}). 
Similarly, $\delta E_{z}=0$ can be obtained as follows. 
From the identity $\nabla_a E^a = 0$ and the field equation 
$\delta E_\mu = 0$ derived above we obtain 
\begin{equation}
0 = \nabla_a \left( g^{ab} \delta E_b \right) 
= \frac{z^{d+1}}{l^2} \partial_z 
\left( z^{-d+1} \delta E_z \right). 
\end{equation}
Therefore, we find 
\begin{equation}
\delta E_z = z^{d-1} C(x), 
\end{equation}
where $C(x)$ is an unknown function of $x^\mu$. 
Using (\ref{deltaga}) and (\ref{hren}) we find 
\begin{align}
C(x) = \left. z^{-(d-1)} \delta E_z \right|_{z=0} 
= - \left. \frac{z^{3-d}}{l^2} \eta^{\mu\nu} 
\partial_\mu \partial_z \delta A_\nu \right|_{z=0}
= - l^{-(d-1)} \delta \VEV{\partial_\mu J^\mu} 
= 0, 
\end{align}
where we have used the conservation of the CFT current 
in (\ref{conservation}). Therefore, we find $\delta E_z = 0$. 
To summarize, we have obtained all the components of 
the linearized field equations $\delta E_{ab}=0$, $\delta E^a=0$ 
from the entanglement first law.

%
\newsection{Antisymmetric tensor field}
The discussion in the previous sections for a vector field 
can be generalized to the case of an antisymmetric 
tensor field. 
To derive the linearized field equation of an antisymmetric tensor 
field we consider a CFT in $d$-dimensional Minkowski spacetime, 
which has an energy-momentum tensor $T_{\mu\nu}$ and an $n$-th rank 
antisymmetric tensor current $J^{\mu_1\cdots\mu_n}$ satisfying 
\begin{equation}
\partial_\mu T^{\mu\nu} = 0, \qquad
T_\mu{}^\mu = 0, \qquad
\partial_{\mu_1} J^{\mu_1\cdots\mu_n} = 0. 
\label{conservation2}
\end{equation}
As in section 2 we introduce a density matrix 
\begin{equation}
\rho_B(\mu) = \frac{e^{-H_B 
+ \mu_{i_1 \cdots i_{n-1}} Q_B^{i_1 \cdots i_{n-1}}/(n-1)!}}{\tr_B  
(e^{-H_B + \mu_{i_1 \cdots i_{n-1}} Q_B^{i_1 \cdots i_{n-1}}/(n-1)!})},
\end{equation}
where $Q_{B}^{i_1\cdots i_{n-1}} 
= \int_B d^{d-1}x J^{0i_1\cdots i_{n-1}}$ is the charge operator 
in $B$ and $\mu_{i_1\cdots i_{n-1}}$ is a constant. 
The charged entanglement entropy is defined as in (\ref{cee}). 
It satisfies the entanglement first law 
\begin{equation}
\delta S_{B}(\mu) = \delta \VEV{H_{B}} 
- \frac{1}{(n-1)!} \, \mu_{i_1 \cdots i_{n-1}} 
\delta \VEV{Q_B^{i_1 \cdots i_{n-1}}}.
\label{firstlaw2}
\end{equation}

The $(d+1)$-dimensional bulk theory dual to this CFT consists of 
the gravitational field $g_{ab}$ and an $n$-th rank antisymmetric 
tensor field $A_{a_1\cdots a_n}$. 
The Lagrangian is given by
\begin{equation}
\mathcal{L} = \frac{1}{16\pi} 
\sqrt{-g} \left( R + \frac{d(d-1)}{l^2}  \right)
- \frac{1}{2(n+1)!} \sqrt{-g} 
F_{a_1\cdots a_{n+1}} F^{a_1\cdots a_{n+1}},
\end{equation}
where the field strength is defined as 
\begin{equation}
F_{a_1 \cdots a_{n+1}} = (n+1) \partial_{[a_1}A_{a_2 \cdots a_{n+1}]}.
\label{fieldstrength}
\end{equation}
Under general variations of the fields the Lagrangian changes as 
\begin{equation}
\delta \mathcal{L} 
= \frac{1}{16\pi} \sqrt{-g} \left( \delta g^{ab} E_{ab} 
+ \frac{16\pi}{n!} \, \delta A_{a_1 \cdots a_n} E^{a_1 \cdots a_n} 
+ \nabla_a v^a \right), 
\end{equation}
where 
\begin{align}
E_{ab} &= R_{ab} - \frac{1}{2} g_{ab} R 
- \frac{d(d-1)}{2l^2} \, g_{ab} - 8\pi T^{\rm bulk}_{ab}, \nonu
E^{a_1 \cdots a_n} &= \nabla_b F^{ba_1 \cdots a_n}, \nonu
T^{\rm bulk}_{ab} &= \frac{1}{n!} \left[ F_{ac_1 \cdots c_n} 
F_b{}^{c_1 \cdots c_n} 
- \frac{1}{2(n+1)} \, g_{ab} F^{2} \right], \nonu
v^a &= \nabla^b \delta g_{ab} - g^{cd} \nabla_a \delta g_{cd} 
- \frac{16\pi}{n!} \, F_a{}^{c_1 \cdots c_n} \delta A_{c_1 \cdots c_n}.
\label{eetv2}
\end{align}
$E_{ab}=0$ and $E^{a_1 \cdots a_n}=0$ are the field equations 
of $g_{ab}$ and $A_{a_1 \cdots a_n}$ respectively with 
$T^{\rm bulk}_{ab}$ being the energy-momentum tensor. 
Under general coordinate transformations 
and antisymmetric tensor gauge transformations
\begin{align}
\delta_{\xi} g_{ab} &= \nabla_a \xi_b + \nabla_b \xi_a, \nonu
\delta_{\xi} A_{a_1\cdots a_n} 
&= \xi^b \partial_b A_{a_1\cdots a_n} 
+ n \partial_{[a_n} \xi^b A_{a_1\cdots a_{n-1}]b} 
+ n \partial_{[a_1} \xi_{a_2\cdots a_n]} \nonu
&= \xi^b F_{ba_1\cdots a_n} + n \partial_{[a_1} \left( 
\xi \cdot A + \xi \right)_{a_2\cdots a_n]},
\label{localtransf2}
\end{align}
the Lagrangian is invariant up to a total divergence as in 
(\ref{localtransfl}).

We split the fields into a background and small perturbations 
around the background: $g_{ab} \rightarrow g_{ab} + \delta g_{ab}$, 
$A_{a_1\cdots a_n} \rightarrow A_{a_1\cdots a_n} 
+ \delta A_{a_1\cdots a_n}$.  
The background is a solution of the field equations 
$E_{ab}=0$, $E^{a_1 \cdots a_n} = 0$ 
and is given by the AdS metric $g_{ab}$ 
in (\ref{background}) and $A_{a_1\cdots a_n}$ satisfying 
$F_{a_1\cdots a_{n+1}} = 0$. 
This background is invariant under the local transformations 
(\ref{localtransf2}) when $\xi^a$ is the Killing vector 
(\ref{bulkkilling}) and the gauge transformation parameter is 
\begin{equation}
\xi_{a_1 \cdots a_{n-1}} 
= \mu_{a_1 \cdots a_{n-1}} - \xi^b A_{ba_1 \cdots a_{n-1}},
\end{equation}
where $\mu_{a_1\cdots a_{n-1}}$ is a constant. This constant 
will be identified with $\mu$ in the first law (\ref{firstlaw2}). 
Since only the space components $\mu_{i_1\cdots i_{n-1}}$ 
transverse to the region $B$ appear in (\ref{firstlaw2}), 
we set other components to zero.

The $(d-1)$-form $\boldsymbol{\chi}$ which satisfies the analog 
of (\ref{eq01}) is given by the same form as (\ref{chiform}), 
where $v^a$ is now given in (\ref{eetv2}) and $\boldsymbol{Q}$ is 
\begin{equation}
\boldsymbol{Q} = \left[ - \frac{1}{16\pi} \nabla^{b} \xi^{c} 
- \frac{1}{2(n-1)!} \, \mu_{a_1\cdots a_{n-1}} 
F^{a_1\cdots a_{n-1} bc} \right] \boldsymbol{\epsilon}_{bc}. 
\label{exact2}
\end{equation}
We impose a gauge condition on the perturbations of the fields as 
\begin{equation}
\delta g_{zz} = 0, \quad
\delta g_{z\mu} = 0, \quad
\delta A_{z\mu_1\cdots \mu_{n-1}} = 0. 
\end{equation}
Integrating $\boldsymbol{\chi}$ over $B$ we obtain 
\begin{align}
\int_B \boldsymbol{\chi} 
&= \int d^{d-1}x \biggl[ \frac{d l^{d-3}}{16R} 
\left\{ R^2 - (x^i-x^i_0)^2 \right\} \delta g_{00}^{(d-2)} \nonu
& \quad + \frac{d-2n}{(n-1)!} \, l^{d-2n-1} \mu^{i_1 \cdots i_{n-1}} 
\delta A^{(d-2n)}_{0 \, i_1 \cdots i_{n-1}} \biggr] \nonu
&= \delta \VEV{H_B} - \frac{1}{(n-1)!} \, \mu_{i_1 \cdots i_{n-1}} 
\delta \VEV{Q_B^{i_1\cdots  i_{n-1}}},  
\end{align}
where we have defined 
\begin{equation}
\delta A_{\mu_1\cdots\mu_n} 
= z^{d-2n} \delta A^{(d-2n)}_{\mu_1\cdots\mu_n}
\label{deltaa}
\end{equation}
for $z \rightarrow 0$ and used the result of the holographic 
renormalization 
\begin{equation}
\delta \VEV{J_{\mu_1\cdots\mu_n}} 
= (d-2n) \, l^{d-2n-1} 
\delta A^{(d-2n)}_{\mu_1\cdots\mu_n}
\label{hren2}
\end{equation}
discussed in appendix A. 
On the other hand, integrating $\boldsymbol{\chi}$ over $\tilde{B}$ 
we obtain 
\begin{equation}
\int_{\tilde{B}} \boldsymbol{\chi} = \delta S_B(\mu), 
\end{equation}
where 
\begin{equation}
S_B(\mu) = \frac{1}{4} A 
- \int_{\tilde{B}} \boldsymbol{\mu} \wedge *\boldsymbol{F}. 
\label{cee3}
\end{equation}
We assume that this $S_B(\mu)$ corresponds to the charged 
entanglement entropy in (\ref{firstlaw2}). 
It contains a term depending on the antisymmetric tensor field 
in addition to the Ryu--Takayanagi formula $\frac{1}{4}A$. 
As in the case of the vector field the relation (\ref{hren2}) 
can be derived also from the entanglement first law 
(\ref{firstlaw2}) and the formula (\ref{cee3}) by considering 
a small size limit of the ball-shaped region $B$ as discussed 
in appendix B.

The entanglement first law (\ref{firstlaw2}) requires (\ref{efl2}) 
with this $\boldsymbol{\chi}$. 
The exterior derivative of $\boldsymbol{\chi}$ is found to be 
\begin{equation}
d \boldsymbol{\chi} 
= \left[ - \frac{1}{8\pi} \delta E^{ab} \xi_b  
+ \frac{1}{(n-1)!} \, \mu_{b_1\cdots b_{n-1}} 
\delta E^{ab_1 \cdots b_{n-1}} \right] \boldsymbol{\epsilon}_a, 
\end{equation}
where $\delta E^{ab}$ is given in (\ref{lfeq}) and 
\begin{equation}
\delta E^{a_1 \cdots a_n} 
= g^{a_1b_1} \cdots g^{a_nb_n} \nabla^c \delta F_{cb_1\cdots b_n}. 
\end{equation}
$\delta E^{ab}=0$ and $\delta E^{a_1 \cdots a_n}=0$ 
are the linearized field equations. 
By requiring (\ref{efl2}) for arbitrary $R$, $x^i_0$, $t_0$ and 
$\mu_{a_1\cdots a_{n-1}}$ in any frame of reference we obtain 
$d$-dimensional components of the linearized field equations 
$\delta E^{\mu\nu}=0$, $\delta E^{\mu_1\cdots\mu_n}=0$. 
Furthermore, the remaining components $\delta E_{z\mu}=0$, 
$\delta E_{zz}=0$, $\delta E_{z\mu_1\cdots \mu_{n-1}}=0$ 
are obtained from the tracelessness and the conservation 
of the energy-momentum tensor and the current (\ref{conservation2}) 
as in the case of a vector field. 
Indeed, from the identity $\nabla_{a_1} E^{a_1\cdots a_n}=0$ 
and the field equation $\delta E^{\mu_1\cdots\mu_n}=0$ we find 
\begin{equation}
\delta E_{z\mu_1\cdots\mu_{n-1}} 
= z^{d-2n+1} C_{\mu_1\cdots\mu_{n-1}}(x),  
\end{equation}
where $C_{\mu_1\cdots\mu_{n-1}}(x)$ is an unknown function of $x^\mu$. 
Using (\ref{deltaa}) and (\ref{hren2}) we find 
\begin{equation}
C_{\mu_1\cdots\mu_{n-1}}(x) 
= \left. z^{-(d-2n+1)} \delta E_{z\mu_1\cdots\mu_{n-1}} \right|_{z=0}
= - l^{-(d-2n+1)} \delta 
\VEV{\partial_\nu J^\nu{}_{\mu_1\cdots\mu_{n-1}}}
= 0
\end{equation}
and therefore $\delta E_{z\mu_1\cdots \mu_{n-1}}=0$. 
Thus, we have obtained all the components of the linearized 
field equations of $g_{ab}$ and $A_{a_1 \cdots a_n}$.

%
\newsection{Conclusions}
In this paper we have shown that the linearized field equations 
of vector and antisymmetric tensor gauge fields as well as 
the gravitational field can be derived from the entanglement 
first law of a CFT with a conserved current. To rewrite the 
first law in terms of the bulk fields we followed the approach 
of \cite{Faulkner:2013ica} and made use of 
the Noether charges of symmetry transformations. 
We considered the gauge transformations of the vector and 
antisymmetric tensor fields as well as the coordinate
transformation. This allows us to obtain the 
linearized field equations of the gauge fields. 
We found that the bulk representations of the charged entanglement 
entropy (\ref{cee2}), (\ref{cee3}) contain the extra 
terms depending on the gauge fields in addition to 
the Ryu--Takayanagi formula.

The derivations of the original Ryu--Takayanagi formula were given 
in \cite{Lewkowycz:2013nqa,Casini:2011kv}.
It would be interesting to study whether our formulae (\ref{cee2}), 
(\ref{cee3}) also can be derived in a similar manner. 
In \cite{Casini:2011kv} the Ryu--Takayanagi formula was derived 
by using the relation between the entanglement entropy for 
the ball-shaped region and the thermal entropy in the hyperbolic 
cylinder $\mathbb{R} \times H^{d-1}$, which we briefly reviewed 
in section 2. By the AdS/CFT correspondence the thermal entropy 
of the CFT is then related to the black hole entropy in the bulk, 
which turns out to equal to the Ryu--Takayanagi formula. 
In this paper we followed more or less this approach at a 
linearized order in perturbations. However, we have not discussed 
a relation of our entropy formulae to black hole entropies. 
It would be better to clarify this point and to confirm our 
formulae. Another approach \cite{Lewkowycz:2013nqa} to derive the 
Ryu-Takayanagi formula uses a bulk generalization of the replica 
trick. It would also be interesting to check whether this approach 
gives our entropy formulae.

The approach in this paper to derive linearized field equations 
from the entanglement first law may be further generalized to 
other bulk fields related to local symmetries. 
For instance, the field equation of a Rarita--Schwinger field 
may be derived from the entanglement first law by considering 
the local supersymmetry. 
On the other hand, it is not clear how to derive field equations 
of bulk fields such as scalar and spinor fields, which are  
not related to local symmetries. 
This is an open problem to be studied in future.

\begin{appendix}

%
\def\numberbysectiona{\@addtoreset{equation}{section}
\def\theequation{A.\arabic{equation}}}
\numberbysectiona
\setcounter{equation}{0}
\setcounter{subsection}{0}
\setcounter{footnote}{0}

\newsection{Holographic renormalization}
In this appendix we briefly discuss the holographic renormalization 
\cite{deHaro:2000vlm,Skenderis:2002wp} of an $n$-th rank 
antisymmetric tensor field $A_{a_1\cdots a_n}$ in $d+1$ dimensions. 
We will obtain the formula (\ref{hren2}) for the one-point function 
of the CFT current $J_{\mu_1\cdots\mu_n}$ used in the text. 
The case of a vector field (\ref{hren}) can be obtained by setting $n=1$. 
The Lagrangian of the antisymmetric tensor field is 
\begin{equation}
\mathcal{L} = - \frac{1}{2(n+1)!} \sqrt{-g} \, F_{a_1\cdots a_{n+1}} 
F^{a_1\cdots a_{n+1}}, 
\end{equation}
where $F_{a_1\cdots a_{n+1}}$ is the field strength 
(\ref{fieldstrength}) and $g_{ab}$ is the AdS metric 
in (\ref{background}). 
We use the gauge condition $A_{z\mu_1\cdots\mu_{n-1}}=0$.

The solution of the field equation derived from this Lagrangian 
can be expanded for small $z$ as 
\begin{align}
A_{\mu_1\cdots \mu_n}(x,z) 
&= A^{(0)}_{\mu_1\cdots \mu_n}(x)
+ z^2 A^{(2)}_{\mu_1\cdots \mu_n}(x) + \cdots \nonu
& \quad + z^{d-2n} A^{(d-2n)}_{\mu_1\cdots \mu_n}(x) 
+ z^{d-2n} \log z^2 B^{(d-2n)}_{\mu_1\cdots \mu_n}(x) + \cdots, 
\label{epsilonexpansion}
\end{align}
where $B^{(d-2n)}_{\mu_1\cdots \mu_n}=0$ when $d$ is odd. 
The field equation gives relations among the coefficient functions. 
The coefficient functions $A^{(m)}_{\mu_1\cdots \mu_n}$ 
($m < d-2n$) and $B^{(d-2n)}_{\mu_1\cdots \mu_n}$ are 
determined as local functions of 
$A^{(0)}_{\mu_1\cdots\mu_n}$ by the field equation. 
In the AdS/CFT correspondence $A^{(0)}_{\mu_1\cdots\mu_n}$ 
plays a role of the source of the CFT current $J^{\mu_1\cdots\mu_n}$, 
while $A^{(d-2n)}_{\mu_1\cdots\mu_n}$ is related to 
the one-point function of the current and represents a CFT state 
\cite{Balasubramanian:1998sn,Balasubramanian:1998de}.

According to the AdS/CFT correspondence the generating functional 
of connected correlation functions of the CFT current is given by 
the classical action evaluated at the solution satisfying 
the Dirichlet boundary condition 
$A_{\mu_1\cdots \mu_n}(x,z=0) = A^{(0)}_{\mu_1\cdots \mu_n}(x)$ 
\cite{Gubser:1998bc,Witten:1998qj}. 
Since the integral over $z$ in the action is divergent near $z=0$, 
we need to regularize it and subtract divergences. 
We regularize the action integral as
\begin{align}
S_{\textrm{reg}} 
&= \int_{z>\epsilon} dz d^d x \, \mathcal{L} \nonu
&= - \frac{1}{2(n+1)!} \int_{z>\epsilon} dz d^d x \, 
\left( \frac{l}{z} \right)^{d-2n-1} F_{a_1\cdots a_{n+1}} 
F^{a_1\cdots a_{n+1}}, 
\label{regaction}
\end{align}
where $\epsilon$ is a small cut-off parameter. 
Here and in the following the raising and lowering of indices 
are done by the flat metric $\eta_{ab}$. 
By integration by parts and using the field equation we can rewrite 
the regularized action as a $d$-dimensional integral at $z=\epsilon$
\begin{equation}
S_{\textrm{reg}} = \frac{1}{2n!} \left( 
\frac{l}{\epsilon} \right)^{d-2n-1} \int_{z=\epsilon} d^d x \, 
A^{\mu_1\cdots \mu_n} \partial_z A_{\mu_1\cdots \mu_n}.  
\label{regaction2}
\end{equation}
Substituting the expansion (\ref{epsilonexpansion}) into this 
action we find that it contains a finite number of divergent 
terms, which are local functionals of 
$A^{(0)}_{\mu_1\cdots\mu_N}(x,\epsilon)$.
To remove the divergences we introduce a counterterm
\begin{equation}
S_{\textrm{ct}} = \int_{z=\epsilon} d^d x \, \mathcal{L}_{\textrm{ct}}, 
\end{equation}
where $\mathcal{L}_{\textrm{ct}}$ is a local function of 
$A^{(0)}_{\mu_1\cdots\mu_n}(x,\epsilon)$. 
This counterterm is chosen such that the renormalized action 
\begin{equation}
S_{\textrm{ren}} = \lim_{\epsilon\rightarrow\, 0} 
\left( S_{\textrm{reg}} + S_{\textrm{ct}} \right)
\end{equation}
is finite. We note that there is an arbitrariness of adding 
finite terms to the counterterm.

The one-point function of the current is then given by 
\begin{align}
\frac{1}{n!} \VEV{J_{\mu_1\cdots\mu_n}(x)} 
= \frac{\delta S_{\textrm{ren}}}{\delta A^{(0)\mu_1\cdots\mu_n}(x)} 
= \lim_{\epsilon\rightarrow\, 0} \, \frac{\delta (S_{\textrm{sub}}
+S_{\textrm{ct}})}{\delta A^{\mu_1\cdots\mu_n}(x,\epsilon)}. 
\label{currentvev}
\end{align}
Here, we have assumed that the coupling of the gauge field to
the current in the CFT Lagrangian has the normalization 
$\mathcal{L}_{\rm CFT} = \cdots 
+ \frac{1}{n!} A^{(0)}_{\mu_1\cdots\mu_n} J^{\mu_1\cdots\mu_n}$. 
Using the regularized action in the form (\ref{regaction2}) we find 
\begin{align}
\frac{\delta (S_{\textrm{sub}}
+S_{\textrm{ct}})}{\delta A^{\mu_1\cdots\mu_n}(x,\epsilon)} 
&= \frac{1}{n!} \left( \frac{l}{\epsilon} \right)^{d-2n-1} 
\left. \partial_z A_{\mu_1\cdots\mu_n}(x,z) \right|_{z=\epsilon}
+ \frac{\delta S_{\textrm{ct}}}{\delta A^{\mu_1\cdots\mu_n}(x,\epsilon)} 
\nonu
&\xrightarrow[\epsilon\rightarrow\,0]{} \frac{1}{n!} (d-2n) \, 
l^{d-2n-1} A^{(d-2n)}_{\mu_1\cdots\mu_n}(x) 
+ \frac{1}{n!} X_{\mu_1\cdots\mu_n}(A^{(0)}), 
\label{currentvev3}
\end{align}
where $X_{\mu_1\cdots\mu_n}(A^{(0)})$ is a function of 
$A^{(0)}_{\mu_1\cdots \mu_n}$, 
which depends on a renormalization scheme. 
Substituting (\ref{currentvev3}) into (\ref{currentvev}) 
we obtain 
\begin{align}
\VEV{J_{\mu_1\cdots\mu_n}(x)} 
= (d-2n) \, l^{d-2n-1} A^{(d-2n)}_{\mu_1\cdots\mu_n}(x) 
+ X_{\mu_1\cdots\mu_n}(A^{(0)}). 
\label{currentvev2}
\end{align}
Taking a variation of the CFT state corresponds to a variation 
of $A^{(d-2n)}_{\mu_1\cdots \mu_n}$ keeping 
$A^{(0)}_{\mu_1\cdots \mu_n}$ fixed 
\cite{Balasubramanian:1998sn,Balasubramanian:1998de}. 
Thus, we obtain (\ref{hren2}) (and (\ref{hren}) for $n=1$) 
in the text.

%
\def\numberbysectiona{\@addtoreset{equation}{section}
\def\theequation{B.\arabic{equation}}}
\numberbysectiona
\setcounter{equation}{0}
\setcounter{subsection}{0}
\setcounter{footnote}{0}

\newsection{Another derivation of (\ref{hren2})}
In this appendix we derive the relation (\ref{hren2}) 
((\ref{hren}) for the $n=1$ vector field case) for the current 
without using the holographic renormalization calculation 
in appendix A. We derive it from the entanglement first law 
(\ref{firstlaw2}) and the holographic charged entanglement 
entropy (\ref{cee3}) by considering a small size 
limit of the ball-shaped region $B$ as was done for 
the energy-momentum tensor in \cite{Faulkner:2013ica}.

In the small size limit $R \rightarrow 0$ the $\mu$-dependent 
term of the right-hand side of (\ref{firstlaw2}) can be 
calculated as 
\begin{align}
(\mbox{RHS}) 
&= - \frac{1}{(n-1)!} \, \mu_{i_1\cdots i_{n-1}} 
\int_B d^{d-1}x \, \delta \VEV{J^{0i_1\cdots i_{n-1}}(x)} \nonu
&= - \frac{1}{(n-1)!} \, \mu_{i_1\cdots i_{n-1}} 
\delta \VEV{J^{0i_1\cdots i_{n-1}}(x_0)} 
\frac{R^{d-1} \Omega_{d-2}}{d-1}, 
\label{b1}
\end{align}
where $\Omega_{d-2}$ is the volume of a unit sphere $S^{d-2}$. 
We have approximated the current by its value at 
the center $x_0^i$. The last factor $R^{d-1} \Omega_{d-2}/(d-1)$ 
is the volume of the region $B$. 
Using (\ref{cee3}) and the gauge condition 
$A_{z\mu_1\cdots \mu_{n-1}}=0$ the $\mu$-dependent term of 
the left-hand side of (\ref{firstlaw2}) is
\begin{align}
(\mbox{LHS}) 
&= - \frac{1}{2(n-1)!} \int_{\tilde{B}} \mu_{i_1\cdots i_{n-1}} 
\delta F^{ab i_1\cdots i_{n-1}} \boldsymbol{\epsilon}_{ab} \nonu
&= \frac{1}{(n-1)!} \int_B d^{d-1}x \, 
\left( \frac{l}{z} \right)^{d-2n-1} \mu_{i_1\cdots i_{n-1}} \nonu
& \quad \times \left. \left( \delta F_{z0i_1\cdots i_{n-1}} 
+ \frac{(x-x_0)^j}{z} \, \delta F_{j0i_1\cdots i_{n-1}} \right)
\right|_{z=\sqrt{R^2-(x^i-x_0^i)^2}}. 
\label{b2}
\end{align}
In the limit $R \rightarrow 0$ this must have the same 
$R$-dependence as (\ref{b1}). It requires that 
the $z \rightarrow 0$ behavior of the field should be 
\begin{equation}
\delta A_{\mu_1\cdots \mu_n}(x,z) 
\sim z^{d-2n} \, \delta A^{(d-2n)}_{\mu_1\cdots \mu_n}(x) 
\end{equation}
as can be seen by rescaling the coordinates $z$ and 
$(x-x_0)^i$ by $R$ as in \cite{Faulkner:2013ica}. 
Then, in the limit $R \rightarrow 0$ the second term of (\ref{b2}) 
vanishes because of the factor $(x-x_0)^j$ while the first term 
gives 
\begin{equation}
(\mbox{LHS}) 
= \frac{d-2n}{(n-1)!} \, l^{d-2n-1} \mu_{i_1\cdots i_{n-1}} 
\delta A^{(d-2n)}_{0i_1\cdots i_{n-1}}(x_0) 
\frac{R^{d-1} \Omega_{d-2}}{d-1},  
\label{b4}
\end{equation}
which has the same $R$-dependence as (\ref{b1}). 
Equating (\ref{b4}) to (\ref{b1}) we obtain (\ref{hren2}).

\end{appendix}


\end{document}